\documentclass[aps,twocolumn,superscriptaddress]{revtex4}
\usepackage{amsmath}
\usepackage{graphicx}

\begin{document}
\bibliographystyle{apsrev}

\title{Decoherence in Nearly-Isolated Quantum Dots}

\author{J. A. Folk}
\affiliation{Department of Physics, Stanford University, Stanford,
California 94305}
\author{C. M. Marcus}
\affiliation{Department of Physics, Stanford University, Stanford,
California 94305}
\affiliation{Department of Physics, Harvard University, Cambridge,
Massachusetts
02138}
\author{J. S. Harris, Jr.}
\affiliation{Department of Electrical Engineering, Stanford University,
Stanford, California 94305}


\begin{abstract} Decoherence in nearly-isolated $GaAs$ quantum dots is
investigated using the change in
average Coulomb blockade peak height
upon breaking time-reversal symmetry. The normalized change in
average peak height approaches the
predicted universal value of $1/4$ at temperatures well below the
single-particle level spacing,
$T < \Delta$, but is greatly suppressed for $T > \Delta$, suggesting 
that inelastic
scattering or other dephasing mechanisms dominate in this regime.
\end{abstract}

\pacs{73.23.Hk, 73.20.Fz, 73.50.Gr, 73.23.-b}
\maketitle

The study of quantum coherence in small electronic systems has been the
subject of intense theoretical and experimental attention
in the last few years, motivated both by questions of fundamental
scientific interest concerning sources of decoherence in materials 
\cite{Mohanty97,
Pierre00, AGA, HuibersDephasing, SIA}, as well as by the possibility 
of using solid state
electronic devices to store and manipulate quantum information \cite{Loss98,
Burkard99}.

Taking advantage of quantum
coherence in the solid state requires a means of isolating the
device from various sources of decoherence, including coupling to
electronic reservoirs. In this context, we have 
investigated
 coherent electron transport through quantum dots weakly
coupled to reservoirs via tunneling
point-contact leads. In this nearly-isolated regime, it is expected
theoretically that inelastic relaxation due to e-e interactions will
vanish below a temperature that is parametrically larger than
the mean quantum level spacing in the dot, $\Delta$
\cite{Altshuler97, Silvestrov97, Leyronas98}. 

It is not obvious,
however, how to measure coherence in nearly-isolated electronic
structures. In this Letter, we introduce a novel method, applicable in
this regime, that uses the change in average Coulomb blockade (CB) peak
height upon breaking time-reversal symmetry as the metric of quantum
coherence within the dot.  By comparing our data to a model of CB transport that
includes both elastic  and inelastic transport processes
\cite{Beenakker00}, we find inelastic rates that are
consistent with dephasing rates
$\tau_{\varphi}^{-1}$  in open quantum dots measured using ballistic 
weak localization
\cite{HuibersDephasing}. Extracting precise values for inelastic
scattering rates using this method appears possible, but would
require a quantitative theory of the crossover from elastic to inelastic
tunneling \cite{Held01}.

When a quantum dot is connected to reservoirs 
(labeled 1,2) via leads with weak tunneling conductance,
$g_{1,2} \ll 1$ (in units of
$e^2/h$), transport is dominated by Coulomb blockade,
which suppresses conduction except at specific voltages
on a nearby gate.  The result is a series of
evenly-spaced, narrow
conduction peaks as the gate voltage is swept, as seen in Fig.\ 1. In this
regime, the usual
techniques for extracting electron decoherence from transport measurements, for
instance using weak localization
\cite{AltshulerMeso, Bergmann84}, are not applicable. Instead, we
take advantage of an analog of weak
localization that reflects a
sensitivity of the spatial statistics of  wave functions to the
breaking of time-reversal symmetry. As in
conventional weak
localization, this effect
changes the {\em average} conductance---or in the
present context, the {\em average} CB peak
height---upon breaking time-reversal symmetry with a weak magnetic
field \cite{ Beenakker00, Alhassid98}.

\begin{figure}
          \label{fig1}
           \includegraphics[width=3.25in]{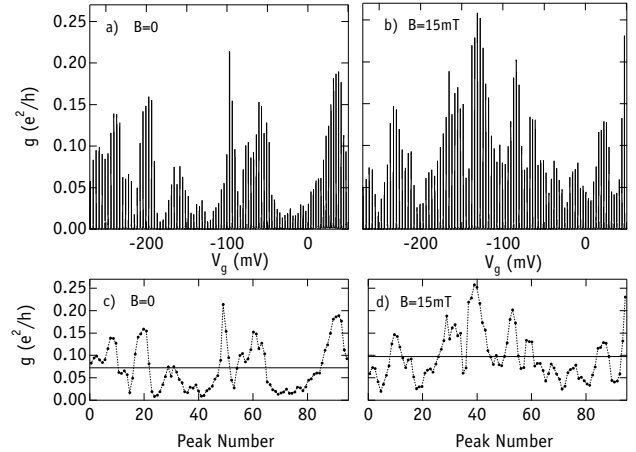}
	\caption{\footnotesize {Coulomb blockade peaks at electron temperature
$T_e = 45\,mK$, for the $0.7 \, \mu
m^2$ device at (a) $B=0$ and (b)  $B=15\,mT$. Every second peak was
measured, as peak-to-peak correlations made measuring each peak
inefficient. (c,d) Peak heights, extracted from (a,b).
Horizontal lines show average peak height, indicating
suppression of average height at $B=0$.}}
\end{figure}

At low temperatures, CB peak heights fluctuate considerably, as seen 
in Fig.\ 1,
reflecting a distribution of tunneling strengths between the quantum 
modes in the
dot and the
leads. When $\Gamma_1,\Gamma_2\ll kT
\ll \Delta$, where $\Gamma_{1(2)} =g_{1(2)} \Delta/ 2\pi$ are the couplings to
the leads, transport occurs via a single eigenstate of the dot. In
this case, CB peaks
are thermally broadened and have a height $g_o=(\pi/2kT)(\Gamma_1
\Gamma_2/(\Gamma_1+\Gamma_2))$ \cite{Alhassid00}. For chaotic or
disordered dots,
universal spatial
statistics of wave functions allow full distributions of CB peak
heights to be
calculated for
both broken ($B\ne0$) and unbroken ($B=0$) time-reversal symmetry
\cite{Alhassid00, JSA}. These distributions have been observed
experimentally \cite{Chang96, Folk96}, with good agreement between theory 
and experiment.

Although not
       emphasized in these
earlier papers, it is readily seen that the two distributions have different
averages. Introducing a dimensionless
peak height $\alpha = (1/\langle\Gamma\rangle)(\Gamma_1
\Gamma_2/(\Gamma_1+\Gamma_2))$ and assuming equivalent leads,
$\langle\Gamma_2\rangle =
\langle\Gamma_1\rangle \equiv\langle\Gamma\rangle$, one finds
$\langle \alpha \rangle_{B=0} = 1/4$ and $\langle \alpha
\rangle_{B\ne 0} = 1/3$. The resulting
difference in average CB peak heights for the two distributions,
normalized by the
average peak height at $B\ne0$,
\begin{eqnarray}
\delta \tilde{g_o}=\delta g_o/\langle g_o \rangle_{B\neq0}= \frac{\langle g_o
\rangle_{B\neq0}-\langle g_o
\rangle_{B=0}} {\langle g_o \rangle_{B\neq0}},
\end{eqnarray}
is then given by $\delta \tilde{g_o} = (\langle \alpha \rangle_{B\ne
0}-\langle \alpha
\rangle_{B=0})/\langle \alpha \rangle_{B\ne 0}= 1/4$. While the
peak heights
themselves are explicitly temperature dependent, this {\em normalized}
difference, $\delta \tilde{g_o}$, does not depend on temperature in the
absence of inelastic processes
\cite{Beenakker00, Alhassid98}.

The absence of explicit temperature dependence of $\delta \tilde{g_o}$ is
not limited to the regime
$kT \ll \Delta$.  As long as transport through the dot is dominated
by elastic scattering ($\Gamma_{el}
\gg \Gamma_{in}$, where $\Gamma_{el}=(\Gamma_1 + \Gamma_2$) is the
broadening due to escape  and
$\Gamma_{in}$ includes all inelastic processes), the normalized
difference in averages does not change
even for
$kT
\gg
\Delta$, i.e., the result
$\delta
\tilde{g_o} = 1/4$ is not affected by thermal averaging. This remains
valid as long as $kT <
(E_{th}, E_c)$, where $E_{th}\sim \hbar/\tau_{cross}$ is the Thouless
energy (inverse crossing time), and
$E_c$ is the charging energy of the dot.

As discussed in Ref.\ \cite{Beenakker00}, the result $\delta
\tilde{g_o} = 1/4$ is
reduced when inelastic processes dominate transport.  In particular, when
$\Gamma_{el}
\ll \Gamma_{in}$, $\delta
\tilde{g_o}(T) \rightarrow 0$ for $kT_e/\Delta \rightarrow \infty$ (see
Fig.\ 2(b)).  The difference in
temperature dependence of $\delta
\tilde{g_o}$ between
$\Gamma_{el}
\ll \Gamma_{in}$ and $\Gamma_{el}
\gg \Gamma_{in}$
arises because for inelastic
transport,
$\langle g_o
\rangle \propto \langle \Gamma_1\rangle
\langle \Gamma_2\rangle/(\langle \Gamma_1\rangle+\langle
\Gamma_2\rangle) $ (the $\Gamma$'s are
averaged individually), whereas  for elastic
transport, $\langle g_o
\rangle \propto \langle \Gamma_1
\Gamma_2/(\Gamma_1+\Gamma_2) \rangle $ (the entire fraction is averaged)
\cite{Beenakker00}. It is this difference in behavior  of $\delta
\tilde{g_o}(T)$ that we use to characterize the relative strength of inelastic
processes.

Previous experiments investigating inelastic broadening of levels in
nearly-isolated quantum
dots have focussed on relaxation of excited states, identifying a
transition from
a discrete to a continuous level spectrum at
$\epsilon>E_{th}$ \cite{Sivan94, Ralph97,
Davidovic99}. Other experiments using coupled quantum dots have
investigated  phonon-mediated
inelastic scattering between dots \cite{Fujisawa98}.  To our 
knowledge, the only
experiment that has addressed the coherence of ground state transport in a single,
nearly-isolated dot (i.e., at low bias,
$eV_{bias} <
\Delta$) are the experiments of Yacoby {\it et al} \cite{Yacoby95} based on
interference in an Aharonov-Bohm ring with a dot ($N
\sim 200$; $\Delta \sim 40\, \mu eV$) in one arm.
Because interference  around
the ring was
observed, the authors inferred a value
$\tau_{\varphi}>10\, ns$ by assuming that the electron coherence time
must be no shorter than the
dwell time of an electron in the dot. This range for
$\tau_{\varphi}$ is somewhat longer than the values measured in open
quantum  dots using ballistic weak localization \cite{HuibersDephasing},
suggesting that some enhancement of $\tau_{\varphi}$ due to
confinement may be occurring in the Yacoby experiment.
However, since dot-in-ring measurements are rather different
from weak-localization
measurements, a direct comparison of
values obtained in the two experiments may not be appropriate.

\begin{table}
         \label{table1}
			\begin{tabular}{cccccc}
			\rule[-2mm]{0mm}{6mm} \hspace{0.1cm} Area
\hspace{0.1cm}
&
$\Delta$ \hspace{0.1cm}
&\hspace{0.4cm} N
&\hspace{0.4cm} $E_{th}$
&\hspace{0.4cm} $E_{c}$
&\hspace{0.4cm} $\epsilon^{**}$
\hspace{0.4cm}
\\
\hspace{0.2cm}$(\, \mu m^2)$
&\hspace{0.2cm} $(\mu eV)$
&
& \hspace{0.4cm}$(\mu eV)$
& \hspace{0.4cm}$(\mu eV)$
& \hspace{0.3cm}$(\mu eV)$
\\
\hline\hline
			0.25 & 28 & 400 & 250& 400& 75 \\
			0.7 & 10 & 1400 &  150& 290&32 \\
			3 & 2.4 & 6000 &  75& 110&10 \\
			8 & 0.9 & 16000 &  45& 65& 5 \\
			\end{tabular}
           \caption{\footnotesize{Device parameters for the four quantum
dots measured:  dot area, $A$, assuming $ 100\,nm$ depletion at
edges; mean spacing of
spin-degenerate levels, $\Delta = 2\pi\hbar^2/m^*A$, where
$m^*$ is the effective mass;  number of electrons in the dot,
$N \sim nA$, where
$n = 2\times10^{11}\, cm^{-2}$ is the 2DEG density; Thouless energy,
$E_{th}$; charging energy
$E_{ch}$; and energy
$\epsilon^{**}$ below which dephasing times due to e-e interactions are
predicted to diverge (see text).}}
\end{table}

We report measurements for four different sized quantum dots
formed in a two-dimensional
electron gas (2DEG), defined using Cr-Au lateral depletion gates on
the surface of a $GaAs/AlGaAs$
heterostructure (see Table I). All dots were made from the same
wafer, which has the 2DEG interface
$90\, nm$ below the surface and a Si doping layer $40\, nm$ from the
2DEG. The electron
density $\sim 2.0\times 10^{11}\, cm^{-2}$ and bulk mobility
$\sim1.4\times10^5\,  cm^2/Vs$ yield a transport mean free path
$\sim1.5\, \mu m$, larger than
or comparable to the lithographic dimensions of the dots, making
transport predominantly ballistic
within the devices.
Measurements were performed
in a dilution refrigerator with base mixing chamber temperature of $25\, mK$.
Electron temperature, $T_e$, in the reservoirs was measured directly using the
width of CB peaks \cite{KouwenhovenRev}, indicating
$T_e = 45\, mK$ at base temperature.

CB peak heights  were
measured by sweeping one of the gate voltages, $V_g$, over
many peaks while simultaneously trimming the gate voltages that control
lead conductances to maintain a constant average transmission with
balanced leads throughout the sweep. This allowed the collection of
$\sim50$ peaks in the smallest dot, and hundreds of peaks in larger
dots (see Fig.\ 1).  Additional ensembles were then collected by making
small changes to the dot shape using other gates. Average peak heights,
$\langle g_o \rangle$, were extracted from these data, collected as a
function of perpendicular magnetic field and normalized by their averages
away from
$B=0$. Figure 2(a) shows that the
functional form for the normalized average peak height,
$\langle \tilde{g_o}(B) \rangle =\langle  g_o(B) \rangle/\langle
g_o\rangle_{B\neq0}$, calculated
within random matrix theory \cite{Alhassid98} agrees well with the
experimental values.    Note
that the average peak height itself, dependent on temperature as well
as the average lead transmissions, cannot be inferred from these
normalized plots.   $\langle
\tilde{g_o}(B) \rangle$ was measured at several temperatures in
each device,
and
$\delta
\tilde{g_o}(T_e)$ was extracted for each.  These are presented in Fig.\ 2(b),
together with the predicted temperature dependences for $\delta
\tilde{g_o}(T_e)$ when either elastic or inelastic transport dominate
\cite{Beenakker00}. Except where otherwise noted, the point contacts were set to give
$\langle g_o\rangle_{B\ne0} \sim 0.05$, though different dot shapes had
average peak height that varied by up to 50\%. The data in Fig. 2(b) 
represent averages over several ensembles at each temperature.

\begin{figure}
          \label{Fig2_cec.EPS}
          \includegraphics[width=2.9in]{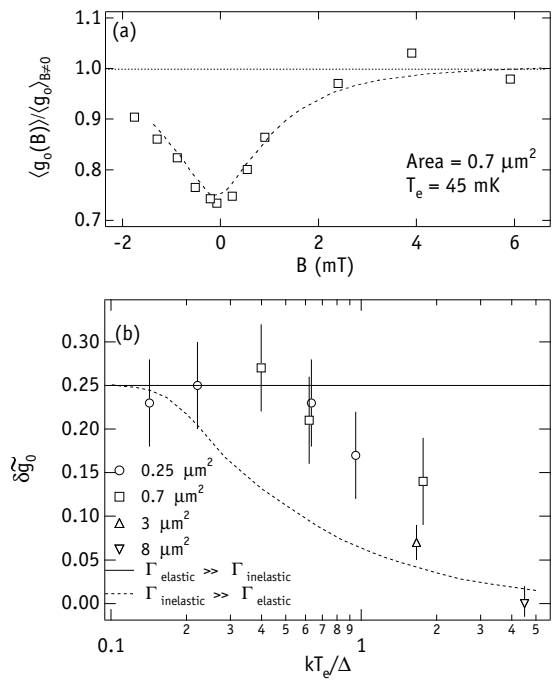}
	\caption{\small{(a) Average peak height as a function of
perpendicular magnetic field,
normalized by the average at $B \ne 0$, for the $0.7 \, \mu
m^2$ dot at $T_e=45\,mK$. Theoretical curve  (dashed) has one adjustable parameter, setting
its width \cite{Alhassid98}. (b) Normalized change in average peak height at $B=0$, $\delta 
\tilde{g_o}$, at several
temperatures,
$T_e$, for all dots measured, along with theoretical values of $\delta
\tilde{g_o}$ when either    elastic
(solid curve) or inelastic (dashed
curve) transport dominate \cite{Beenakker00}.  Note crossover from solid 
to dashed curve
around
$kT_e/\Delta\sim 1$.}}
\end{figure}

In the $0.25\,\mu m^2$ dot at
$T_e =45\,mK$ and $70\,mK$,
$\delta \tilde{g_o}$ was consistent with $1/4$ as expected since $kT_e
\ll
\Delta$ for both temperatures.  In this regime, one cannot distinguish
between elastic and
inelastic scattering since both mechanisms give $\delta \tilde{g_o}
\simeq 1/4$.  In the
$0.7
\mu m^2$ device at
$45\,mK$, we again find $\delta
\tilde{g_o}\sim 0.25$. In this dot, however, $45\, mK$ corresponds to
$kT_e/\Delta \sim 0.5$.
For $\Gamma_{in}\gg \Gamma_{el}$, a ratio $kT_e/\Delta \sim 0.5$
gives a predicted value
for the average peak height difference of $\delta
\tilde{g_o} \sim 0.13$ (see the dashed curve in Fig.\ 2(b))
whereas for
$\Gamma_{el} \gg \Gamma_{in}$, $\delta
\tilde{g_o} = 0.25$ for all values of $kT_e/\Delta$ (solid line in Fig.\
2(b)).  We therefore conclude that
$\Gamma_{in}<\Gamma_{el}$ in the
$0.7 \mu m^2$  device at $45\,mK$, when the point
contact transmissions are set so that $\langle
g_o\rangle \sim 0.05$.  We can extract $\Gamma_{el}$ from
average peak height $\langle
g_o\rangle$ using the equation
$\Gamma_{el}\sim \langle g_o\rangle\Delta$, valid in the regime
$kT_e\gtrsim\Delta$ \cite{Alhassid00}.  For $\langle
g_o\rangle \sim 0.05$ in the $0.7\,\mu m^2$ device, this gives
$\Gamma_{el} \sim 0.5\,  \mu eV$,
and we therefore conclude $\Gamma_{in} < 0.5\, \mu eV$ at $45\, mK$.

\begin{figure}
          \label{Fig3_cec.EPS}
          \includegraphics[width=3in]{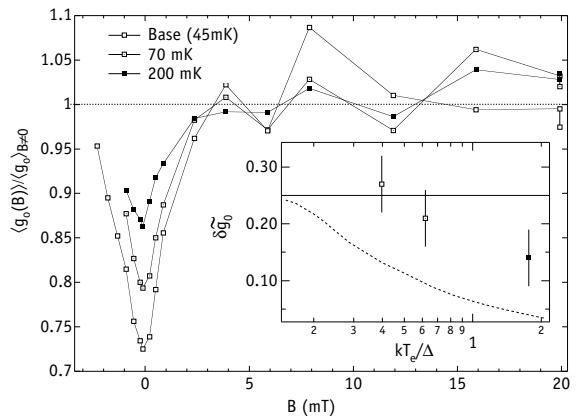}
	\caption{\small{(a) Normalized average peak height as 
a function of
perpendicular magnetic field, for the $0.7 \, \mu
m^2$ dot at several temperatures. Inset shows $\delta
\tilde{g_o}$ for each temperature, along with theoretical curves from Ref.
\cite{Beenakker00}.  Note crossover from solid to dashed curve at 
$T_e \sim 200mK$.}}
\end{figure}

Similarly, we can observe for each dot (with different values of $\Delta$), at each
temperature, whether transport is principally elastic or
inelastic, or whether the two rates are comparable.  Measurements
of $\langle
\tilde{g_o}(B)\rangle$ in the $0.7 \mu m^2$ device at $45\,mK$,
$70\,mK$, and $200\,mK$ are shown in
Fig.\ 3, with the extracted values of $\delta
\tilde{g_o}(T)$ shown in the inset.  For the
$0.7
\mu m^2$ device, we find that $\Gamma_{el} > \Gamma_{in}$ at
$45\,mK$ and $70\,mK$, whereas by $200\,mK$ the crossover to the lower curve $(\Gamma_{el}
< \Gamma_{in})$ has begun, presumably because
$\Gamma_{in}$ increases at higher temperature.  We
infer that a $0.7\mu m^2$ device at $200mK$ is in the crossover regime
$\Gamma_{in}\sim0.5\mu eV$.

  We observe a similar crossover from
$\Gamma_{el} > \Gamma_{in}$ to $\Gamma_{el} <
\Gamma_{in}$ by changing $\Gamma_{el}$ at
fixed temperature.  Figure 4 shows $\langle
\tilde{g_o}(B)\rangle$ in the $0.7 \mu m^2$ device at
$200\,mK$ for three different settings of the point contacts,
ranging from  $\langle
g_o\rangle_{B \neq 0} = 0.016$ to $\langle
g_o\rangle_{B \neq 0} = 0.057$; the extracted values for
$\delta \tilde{g_o}$ are shown in the inset.  Despite significant
statistical uncertainty, it is clear that
$\delta
\tilde{g_o}$ decreases as $\Gamma_{el}$ decreases.  We note that
in the same device at $45\,mK$ and $70\,mK$ there is no
difference  in $\delta \tilde{g_o}$ for
the same of point contact transmissions, within experimental uncertainty.  This is
presumably because $
\Gamma_{in}$ is lower at these temperatures, and
$\Gamma_{el} > \Gamma_{in}$ for all point contact
transmissions measured.

\begin{figure}
          \label{Fig4_cec.EPS}
          \includegraphics[width=3.2in]{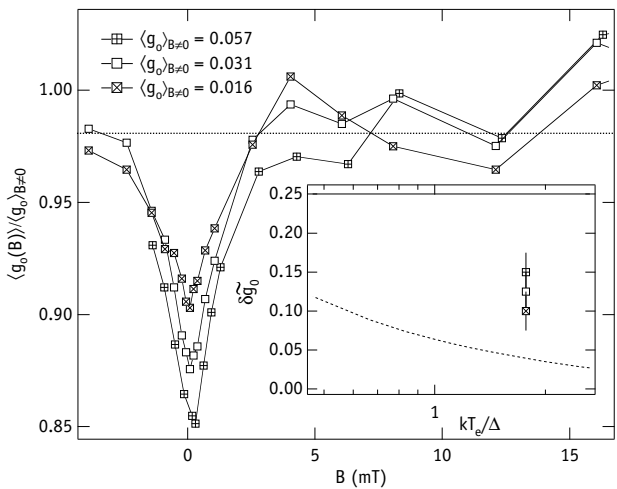}
	\caption{\small{(a) Normalized average peak height as 
a function of
perpendicular magnetic field, for the $0.7 \, \mu
m^2$ dot at $T_e=200mK$ for three settings of the point contacts. Inset
shows
$\delta
\tilde{g_o}$ for setting, along with theoretical curves from Ref.
\cite{Beenakker00}.  As $\Gamma_{in}$ is decreased
by  closing point contacts,
experimental $\delta
\tilde{g_o}$ moves away from solid curve ($\Gamma_{el}>\Gamma_{in}$)
toward dashed curve ($\Gamma_{el}<\Gamma_{in}$), as one would expect.}}
\end{figure}

Based on recent theoretical arguments, one expects inelastic
scattering due to electron-electron interactions  to be strongly
suppressed in isolated
quantum dots for
$kT<\epsilon^{**}$, where $\epsilon^{**} \sim N^{1/4}\Delta$ for
ballistic chaotic dots containing $N$ electrons
\cite{Altshuler97, Silvestrov97, Leyronas98}. Because this suppression 
is not expected
to occur in open dots,
it is useful to compare the constraints on inelastic rates
discussed above for
nearly-isolated dots with experimental values of the phase coherence
time $\tau_{\varphi}$
measured in open dots
\cite{HuibersDephasing}.  Although there may be dephasing
mechanisms that do not involve inelastic processes, the
inelastic scattering rate should provide a lower bound for the dephasing
rate
$\tau_{\varphi}^{-1}$. Dephasing rates extracted from weak localization in
open quantum dots are found to be well described by the empirical relation 
$\hbar/\tau_{\varphi}(T_e)
\sim 0.04\, kT_e$ over the range of temperatures $\sim 70\,mK-300\,mK$,
independent of dot
size \cite{HuibersDephasing}. For the closed dots
we again may use $\Gamma_{el}\sim \langle g_o\rangle\Delta$, giving a
ratio of elastic scattering rate to dephasing rate in the corresponding
open dots 
$\Gamma_{el}/(\hbar/\tau_{\varphi}) \sim (\langle g_o\rangle / 0.04)\,
kT_e/\Delta$. If, for the sake of comparison, we
   identify
$\Gamma_{in}$ with $\hbar/\tau_{\varphi}$, we would then expect for
$\langle g_o\rangle_{B\ne0} \sim 0.05$ a
ratio
$\Gamma_{el}/\Gamma_{in}
\sim kT_e/\Delta$, suggesting a crossover between the
curves in Fig.\ 2(b) for $kT_e/\Delta \sim 1$.  The data in Fig.\ 2(b) do show
a crossover in the vicinity
of $kT_e/\Delta\sim 1$, consistent with the identification
$\Gamma_{in}^{(closed)}\sim
(\hbar/\tau_{\varphi})^{(open)}$. For a more quantitative comparison 
between dephasing in
open dots and inelastic scattering through nearly-isolated dots, one 
would need a
theoretical calculation of
$\delta
\tilde{g_o}$ in the regime
$\Gamma_{el} \sim \Gamma_{in}$ \cite{Held01}.

We do not see evidence for the predicted \cite{Altshuler97, 
Silvestrov97, Leyronas98} divergence of the coherence time
for 
$kT_e/\Delta < N^{1/4}\sim 5$.  A possible explanation is
that electron-electron  interactions are not
the primary dephasing mechanism in our system.  Several other 
mechanisms have been
proposed, including external radiation
\cite{AGA, Vavilov99}, two-level systems
\cite{KondoDephasing}, and nuclear
spins \cite{Dyugaev00}.  We
cannot, however,
rule out some enhancement of coherence due to confinement at
a level reported in
\cite{Yacoby95}.  The lack of a quantitative theory in the crossover regime
$\Gamma_{in}\sim\Gamma_{el}$ prevents us from extracting exact values
for $\Gamma_{in}$
from our data.

In conclusion, we have developed a new method of measuring inelastic rates in a
nearly-isolated quantum dot, using the change in average CB peak height upon
breaking time-reversal symmetry.
These measurements appear consistent
with dephasing times previously measured in open ballistic quantum
dots, however for a
careful quantitative analysis of our data we await a theory treating the regime
$\Gamma_{in}\sim\Gamma_{el}$.
 
We thank S. Patel for device fabrication and C. Duru\"oz for material 
growth. We
acknowledge valuable discussions with I. Aleiner, Y. Alhassid, B. 
Altshuler, C. Beenakker,
P. Brouwer, L. Glazman, and K. Held.  This work was supported in part 
by the ARO under
341-6091-1-MOD 1 and DAAD19-99-1-0252.  JAF acknowledges partial 
support from the DoD.

{

}
\end{document}